\documentclass{aipproc}
\layoutstyle{6x9}
\usepackage{gensymb}
\usepackage{units}

\begin{document}
\title{Open heavy flavour analysis with the ALICE experiment at LHC}
\author{Serhiy Senyukov for ALICE collaboration}{address={Universit$\grave{a}$ degli Studi del Piemonte Orientale, Viale Teresa Michel, 11 15121 Alessandria, Italy}}
\begin{abstract}
The ALICE experiment at the LHC was taking data in proton-proton collisions at the center-of-mass energy of \unit[7]{TeV} starting from March till the end of October 2010. First heavy-ion collisions were delivered during November 2010. Among other particles, charmed and beauty mesons can be reconstructed by the ALICE apparatus. Open heavy flavour mesons provide a powerful tool to study hot quark matter produced in high energy heavy-ion collisions because their spectra are expected to be affected by energy loss in the medium. Measuring such particles is important also in proton-proton collisions. In this case they provide a necessary reference for heavy-ion collisions and allow to test pQCD predictions in a new energy domain. Different ways to reconstruct D and B mesons are described together with adopted selection and analysis techniques. D mesons in the central rapidity region are reconstructed via 2,3, and 4-prongs hadronic decays and via single electrons. In the forward region, semi-muonic decays are used to measure the production of D and B mesons. The first results for D mesons, single electrons and single muons are reported.
\end{abstract}
\maketitle
\section{Introduction}
ALICE (A Large Ion Collider Experiment) \cite{ALJINST} is one of the four main experiments at the LHC. Its main task is to study ultra-relativistic collisions of heavy ions. During these collisions the energy density can reach \unit[10]{ GeV/fm$^3$} and the temperature can exceed \unit[200]{MeV}. According to the lattice QCD predictions a new state of matter called Quark-Gluon Plasma (QGP) can be formed under such conditions \cite{SHU}. Heavy flavour quarks provide a powerful tool for studying the QGP properties. Due to their large mass $c$ and $b$ quarks are mainly produced in hard parton-parton scatterings in the early stages of the collision. Therefore they interact with the medium created. Due to the interaction with the medium, partons lose part of their kinetic energy. The magnitude of this loss is related to the properties of the medium, especially to the energy density. Moreover there are predictions  of a parton mass dependence of the energy loss, also referred to as the "dead cone effect" \cite{Dead-cone_effect}. Energy loss for quarks is expected to be smaller compared to that for gluons. This difference can be attributed to the stronger interaction of gluons with the QGP. More details on energy loss can be found in \cite{Vitev:2008jh}. The experimental observable that is used to describe medium effects is the nuclear modification factor $R_{AA}$ defined as follows:
\begin{equation}\label{Raa_definition}
    R_{AA}(p_T)=\frac{d^2N_{AA}/dp_T dy}{<N_{coll}> \times d^2N_{pp}/dp_T dy}
\end{equation}
where $d^2N_{AA}/dp_T dy$, $d^2N_{pp}/dp_T dy$ are the differential yields in nucleus-nucleus and proton-proton collisions and $N_{coll}$ is the number of binary nucleon-nucleon collisions. In the absence of interaction with matter $R_{AA}$ should be equal to unity. Measurement of the heavy-flavour production in proton-proton collisions is necessary to obtain the nuclear modification factor. Moreover, knowledge of charm and beauty production in pp collisions at LHC allows to test pQCD models in a new energy domain.
\section{ALICE detector}
The ALICE setup consists of two main parts: a central barrel and a forward muon arm.

The central system covers the full azimuthal angle in $|\eta|\leq0.9$ and is embedded in a magnetic field $B\leq \unit[0.5]{T}$. The central detectors used for the open heavy flavour analyses are, in the order from the interaction vertex to the outside, the Inner Tracking System (ITS), the Time Projection Chamber (TPC) and the Time Of Flight (TOF). The ITS detector consists of six layers of silicon detectors. The two innermost layers are made with Silicon Pixel Detectors (SPD)followed by two layers of Silicon Drift Detectors (SDD). The two outermost layers consist of Silicon Strip Detectors (SSD). The ITS is used for the primary and secondary vertex reconstruction. Impact parameter resolution of the ITS is of the order of a few hundred microns for very low momentum tracks. It improves with increasing track momentum and reaches \unit[80]{$\mu$m} at \unit[1]{GeV/c}. The TPC is the main tracking device in ALICE. It provides up to 159 space points per track. The momentum resolution achieved is about 1\% for tracks with $p_t<\unit[1]{GeV/c}$. Information on the specific energy loss dE/dx provided by the TPC is used for particle identification (PID). The TOF detector extends the PID capabilities of ALICE up to \unit[4]{GeV/c}.

Event triggers are based on signals in the SPD and the VZERO detectors. VZERO is composed of two arrays of scintillator counters (VZERO-A and VZERO-C) located on either side of the interaction point.

The muon spectrometer covers the polar angular range $171\degree < \theta < 178\degree$ ($-4.0 < \eta < -2.5$). It is composed of a passive front absorber, a beam shield, a \unit[3]{Tm} dipole magnet, five stations of high granularity tracking chambers, each based on two planes of Cathode Pad Chambers. Finally, two stations of trigger chambers equipped with two planes of Resistive Plate Chambers each are located downstream of the tracking system, after a \unit[1.2]{m} thick iron wall.

More technical details on the design of the ALICE detector can be found in \cite{ALJINST}.
\section{Event samples}
ALICE collected data on proton-proton collisions at $\sqrt{s}=\unit[7]{TeV}$ between March and October 2010. Two main classes of events are used for the open heavy flavour analyses. A minimum bias trigger requires a signal in one of the SPD or VZERO-A or VZERO-C in coincidence with the proton bunches passing through the interaction point. Thus, a single charged particle in eight rapidity units is sufficient to generate this trigger. These events are used for exclusive reconstruction of the D mesons and inclusive reconstruction of single electrons in the central rapidity region. The study of single muons is based on a dedicated event class triggered by a muon track passing the acceptance of the muon spectrometer. During 2010 pp running ALICE recorded $8\times10^8$ minimum bias events and $1\times10^9$ events triggered by the muon spectrometer. Results presented here are based on a subsample containing about $1.4\times10^8$ minimum bias events and $4.7\times10^6$ muon events.
\section{First results on the heavy flavour reconstruction}
Three different channels are used in ALICE for the reconstruction of open heavy flavour particles. D mesons are exclusively reconstructed via hadronic decays in the central barrel. Single electrons from semi-leptonic decays allow to asses open charm and beauty production at the central rapidity region. Open heavy flavour production at forward rapidity is accessible by means of single muons originated from semi-leptonic decays. The feasibility of these analyses was first studied using Monte Carlo simulations. More details can be found in \cite{ALICE_PPR2,DAINESE_PHD,BRUNA_PHD,Senyukov:2008zz,Senyukov_PHD,Rossi_PHD,Romita_PHD}
\subsection{Exclusive reconstruction of D mesons}
Exclusive reconstruction of D mesons is based on the invariant mass analysis of the two-, three- or four-track combinations emerging from a common displaced vertex. The selection of the candidate topologies is performed in three steps. The first step is the quality selection of single tracks. In the second step particle identification is applied in order to reduce the number of possible combinations. Then candidate topologies are formed according to the decay channel under consideration. Different topological and kinematical cuts are applied to reduce the combinatorial background and to improve the statistical significance of the measurement.
\subsubsection{$D^0$ meson}
\begin{figure}
  \includegraphics[width=1.0\textwidth]{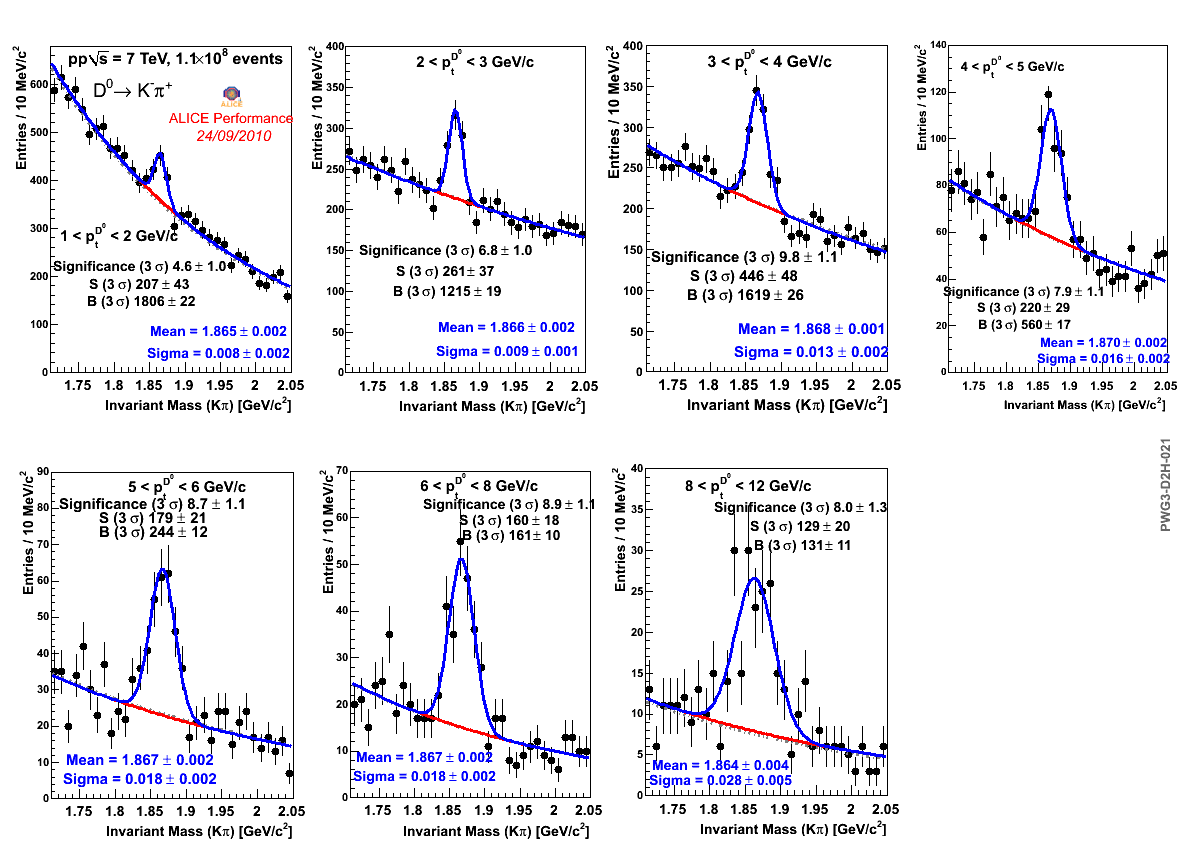}\\
  \caption{Invariant mass spectra of $D^0\rightarrow K^-\pi^+$ candidates at different transverse momenta from $1.1\times10^8$ minimum bias events at $\sqrt{s}=\unit[7]{TeV}$.}\label{fig:D02Kpi_bins}
\end{figure}

The main channel for the $D^0$ meson reconstruction is the two-body decay into a kaon and a pion: $D^0\rightarrow K^-\pi^+$. This channel has a branching ratio of about 4\% and is characterized by a low combinatorial background. The candidate selection is based on the product of the impact parameters of the two tracks ($d_0^K \times d_0^\pi$) and the cosine of the pointing angle ($cos(\theta_{point})$). Here the pointing angle is defined as the angle between the direction of the total momentum and the line connecting the primary and the secondary vertices. Invariant mass spectra of $D^0$ candidates for different $p_T$ bins are shown on Fig. \ref{fig:D02Kpi_bins}. With the statistics available the $D^0$ meson can be reconstructed via the two prong decay in the $p_T$ range from \unit[1]{GeV/c} to \unit[12]{GeV/c}.

\begin{figure}[h]
  \includegraphics[width=1.0\textwidth]{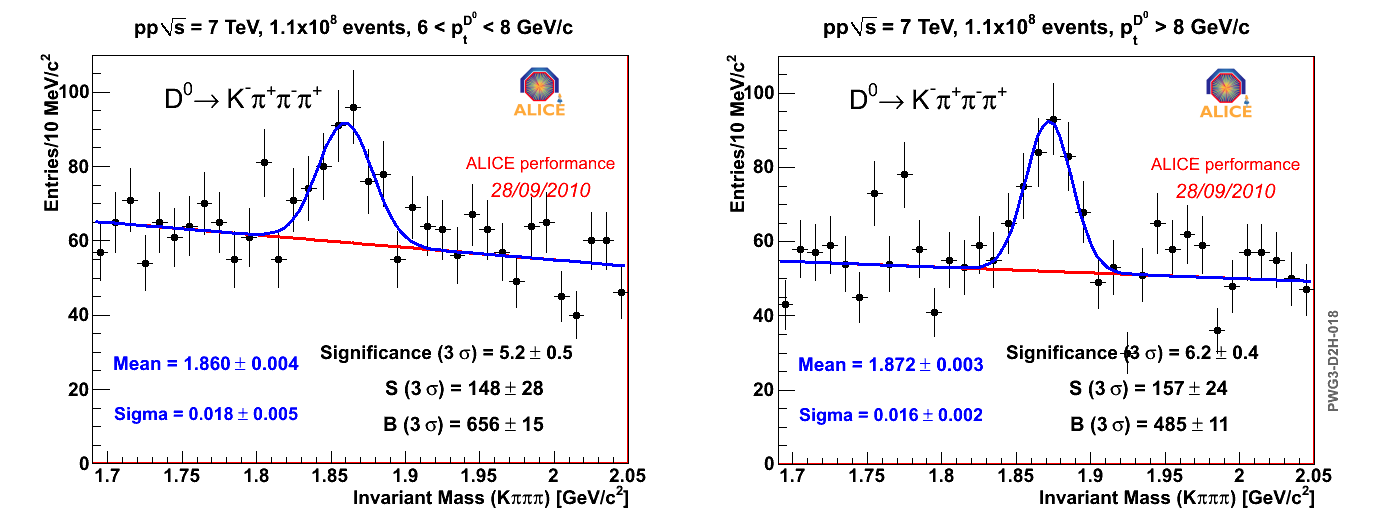}\\
  \caption{Invariant mass spectra of $D^0\rightarrow K^-\pi^+\pi^-\pi^+$ candidates at $6<p_T<\unit[8]{GeV/c}$ (left panel) and $p_T>\unit[8]{GeV/c}$ (right panel) from $1.1\times10^8$ minimum bias pp events at $\sqrt{s}=\unit[7]{TeV}$. }\label{fig:D02Kpipipi}
\end{figure}

The $D^0$ meson can be also reconstructed via its four-body decay into a kaon and three pions: $D^0\rightarrow K^-\pi^+\pi^-\pi^+$ This channel has a larger branching ratio ($BR\simeq8\%$) but also larger combinatorial background. Figure \ref{fig:D02Kpipipi} shows the invariant mass spectra of $D^0$ candidate topologies at $6<p_T<\unit[8]{GeV/c}$ (left panel) and $p_T>\unit[8]{GeV/c}$ (right panel). The position of the mass peak is in agreement with that found in the two-prong decay.

\subsubsection{$D^+$ meson}
\begin{figure}
  \includegraphics[width=1.0\textwidth]{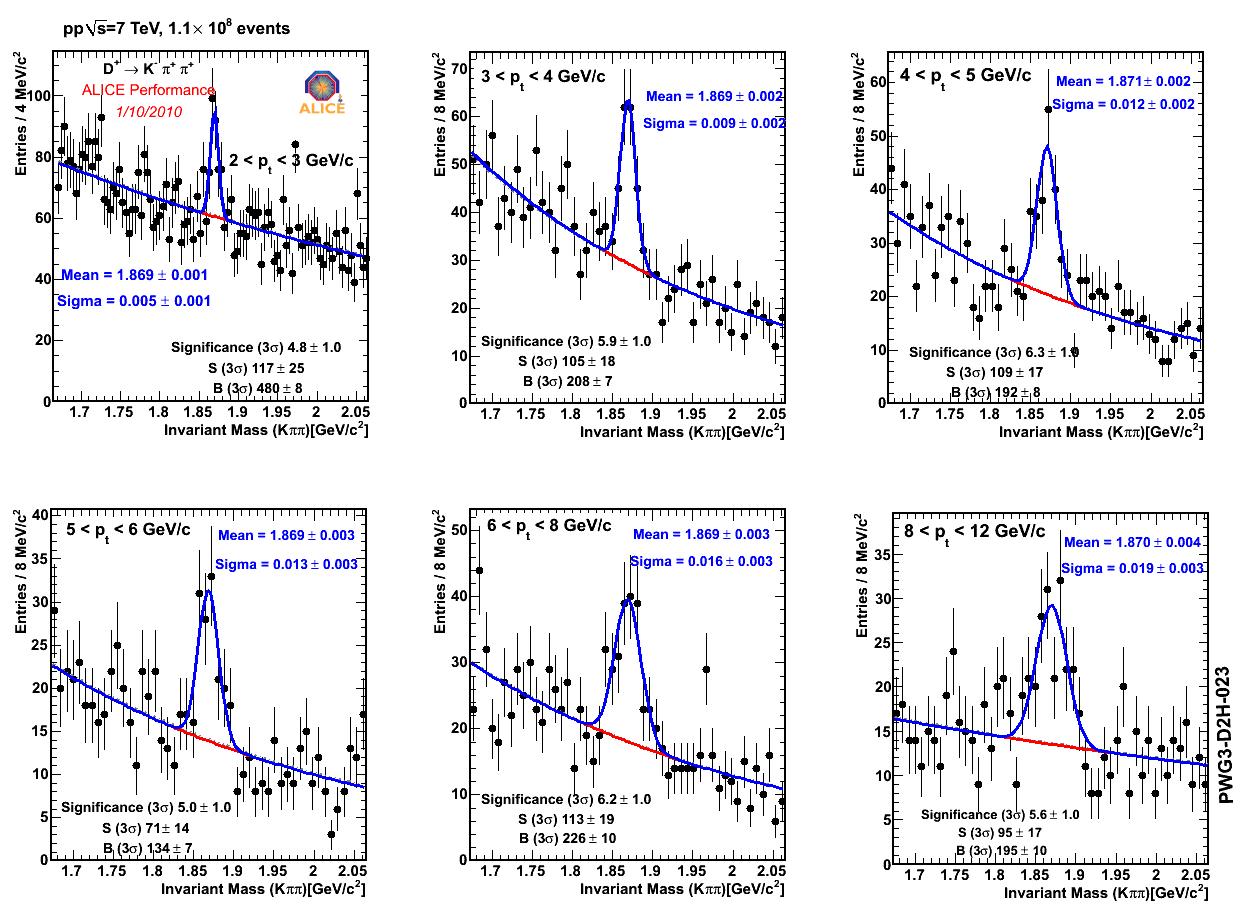}\\
  \caption{Invariant mass spectra of $D^+\rightarrow K^-\pi^+\pi^+$ candidates at different transverse momenta from $1.1\times10^8$ minimum bias pp events at $\sqrt{s}=\unit[7]{TeV}$.}\label{fig:Dplus_bins}
\end{figure}

The $D^+$ meson is reconstructed in the channel $D^+\rightarrow K^-\pi^+\pi^+$. Due to the presence of three daughter particles this channel is affected by a higher combinatorial background than the two-body decay of the $D^0$ meson. On the other hand, the $D^+$ has larger decay length ($c\tau\simeq$ \unit[311]{$\mu$m} compared to \unit[123]{$\mu$m} for the $D^0$ meson ). This makes the selection based on the distance between the primary and the secondary vertex more effective. Figure \ref{fig:Dplus_bins} shows invariant mass spectra of $D^+\rightarrow K^-\pi^+\pi^+$ candidates at different transverse momentum. With the statistics available the $D^+$ meson can be reconstructed in the $p_T$ range from \unit[2]{GeV/c} to \unit[12]{GeV/c}.
\subsubsection{$D^{*+}$ meson}
\begin{figure}
  \includegraphics[width=1.0\textwidth]{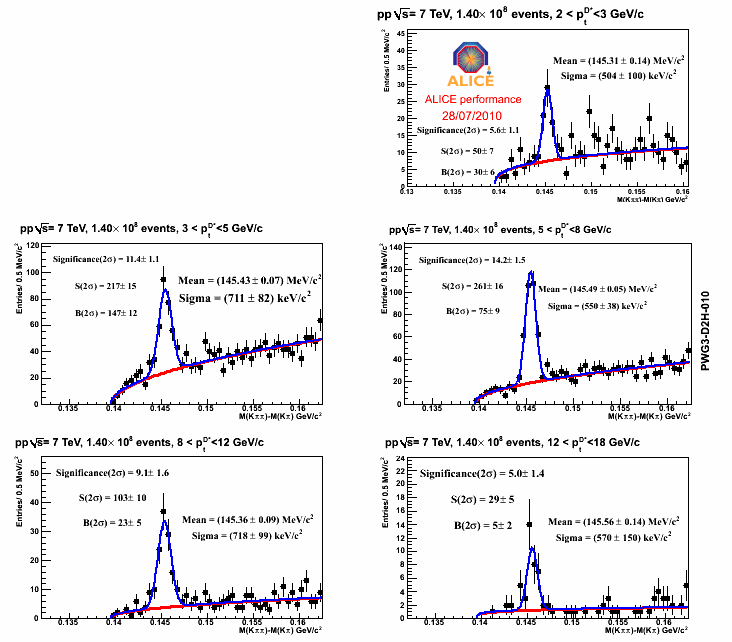}\\
  \caption{$M(K\pi\pi)-M(K\pi)$ invariant mass difference spectra for $D^{*+}$ candidates at different transverse momenta from $1.1\times10^8$ minimum bias pp events at $\sqrt{s}=\unit[7]{TeV}$.}\label{fig:Dstar_bins}
\end{figure}

The $D^{*+}$ meson is reconstructed via the decay channel $D^{*+}\rightarrow D^0\pi_S^+$. This is a strong decay and therefore its secondary vertex is not directly accessible. Thus the daughter $D^0$ candidates should be selected via the channel $D^0\rightarrow K^-\pi^+$. The selection strategy is the same as for $D^0$ but cut values are looser. In the next step the distribution of the following mass difference is build: $M(K\pi\pi)-M(K\pi)$. $D^{*+}$ manifests itself as a peak at the value equal to the mass difference between $D^{*+}$ and $D^0$ (i.e. $\sim$\unit[145]{MeV/c$^2$}). Such distributions for the $D^{*+}$ candidates in different $p_T$-bins are shown in Fig. \ref{fig:Dstar_bins}. With the statistics available the $D^{*+}$ meson reconstruction is possible for transverse momenta up to \unit[18]{GeV/c}.

\subsubsection{$D_s^+$ meson}
\begin{figure}[h]
  \includegraphics[width=0.65\textwidth]{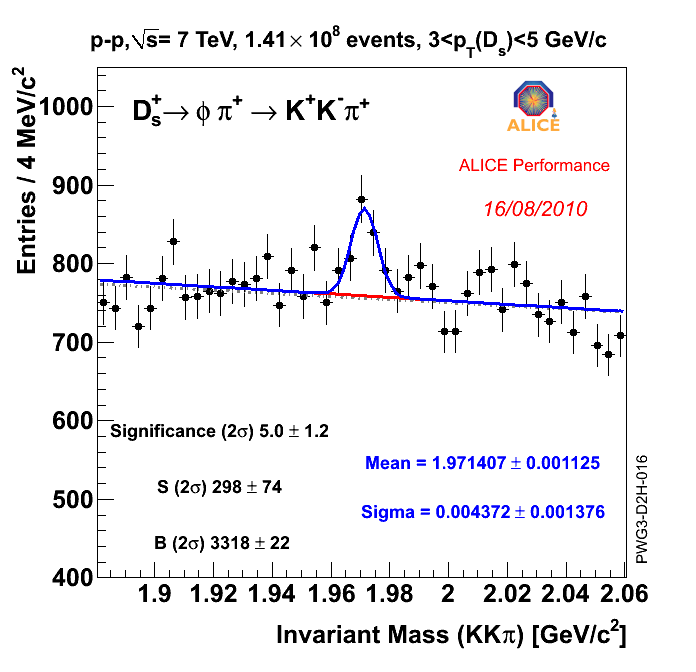}\\
  \caption{$KK\pi$ invariant mass spectrum corresponding to $1.4\times10^8$ minimum bias events at $\sqrt{s}=\unit[7]{TeV}$. $D_s^+$ candidates with $3<p_T<\unit[5]{GeV/c}$ are considered.}\label{fig:Ds_3_5}
\end{figure}

The $D_s^+$ meson is reconstructed in the channel $D_s^+\rightarrow K^+K^-\pi^+$. This decay channel is characterized by a low branching ratio ($BR\simeq5.5 \%$) and short decay length ($c\tau\simeq$ \unit[150]{$\mu$m}). These two factors together with higher combinatorial background makes $D_s^+$ meson reconstruction a challenging task. $D_s^+$ mesons preferentially decay through intermediate resonant states ($\phi$ and $K^{*0}$). This fact is used to improve the background suppression by requiring that the invariant mass of two of the daughter tracks is close to the mass of the $\phi$ or of the $K^{*0}$. Invariant mass spectrum for the $D_s^+\rightarrow K^+K^-\pi^+$ candidates at $3<p^{D_s}_T<\unit[5]{GeV/c}$ is shown on Fig. \ref{fig:Ds_3_5}
\subsection{Single electron analysis}
The reconstruction of single electrons gives another way to asses open charm production at central rapidity. Moreover, it allows to estimate the yield of open beauty. Reliable particle identification (PID) is important for this analysis in order to reject abundant contribution from hadron tracks.

\begin{figure}[h]
  \includegraphics[width=1.0\textwidth]{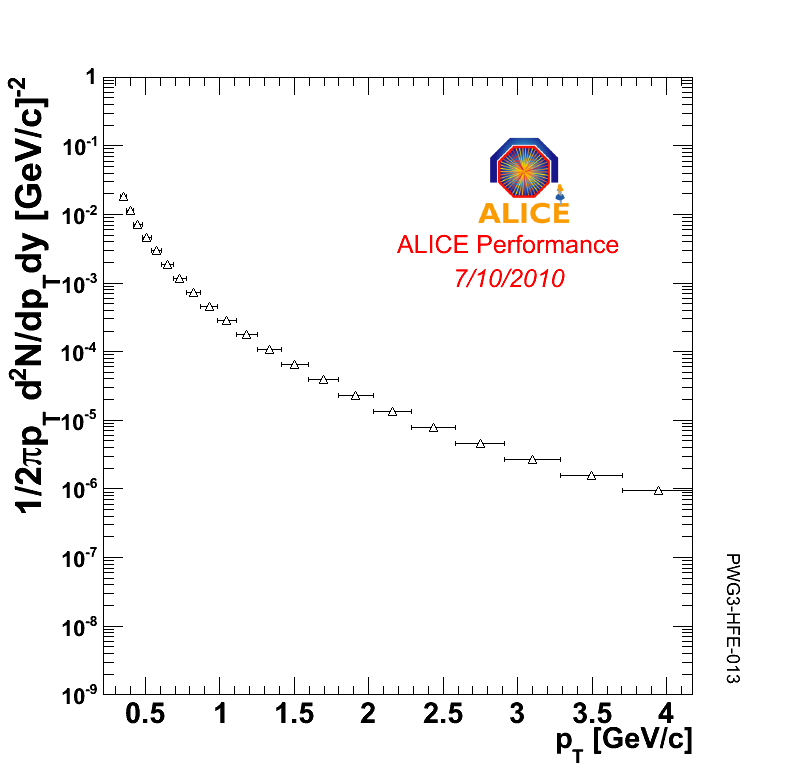}\\
  \caption{Inclusive transverse momentum spectrum of the single electrons obtained from $1.1\times10^8$ minimum bias pp events at $\sqrt{s}=\unit[7]{TeV}$. The spectrum is corrected for efficiencies and acceptance and is unfolded in $p_T$. Residual hadron contamination is subtracted.}\label{pic:single_electrons_pt}
\end{figure}

The current PID strategy is based on electron selection using the dE/dx signal in the TPC. Remaining contaminations from protons and kaons at low $p_T$ are rejected using TOF information. The corrected transverse momentum distribution of the selected tracks is shown in Fig. \ref{pic:single_electrons_pt}. This spectrum is corrected for the acceptance and efficiencies. Residual hadron contamination is subtracted and unfolding in $p_T$ is performed. The Transition Radiation Detector (TRD) will allow to extend electron identification to large $p_T$. At the moment its PID calibration is ongoing.
\subsection{Single muon analysis}
\begin{figure}[h]
  \includegraphics[width=1.0\textwidth]{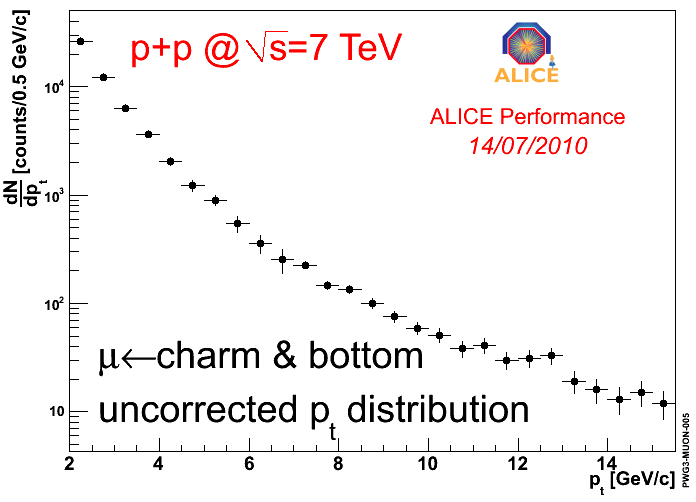}\\
  \caption{Single muon transverse momentum distribution in proton-proton collisions at $\sqrt{s}=\unit[7]{TeV}$.}\label{pic:single_muon_pt}
\end{figure}

Production of open charm and beauty at forward rapidity can be measured via single muons reconstructed in the muon spectrometer. In order to obtain the spectrum of the muons originating from heavy flavour decays several background components must be subtracted. The hadron component of the background is removed by requiring that the track reconstructed in the tracking part of the detector matches the track in the trigger system. The muon contribution from decays of primary kaons and pions is subtracted using Pythia results (tune ATLAS-CSC) normalized to the data in the low-$p_T$ region. The remaining background is due to the muons coming from decays of secondary light hadrons. For $p_T>\unit[2]{GeV/c}$ this component can be neglected. The resulting uncorrected $p_T$ spectrum is shown in Fig. \ref{pic:single_muon_pt}. The present statistics allows to investigate the $p_T$ range up to \unit[15]{GeV/c}.

\section{Conclusions}
The production of open heavy flavour is an important tool to study the QGP formed in PbPb collisions at the LHC. It is crucial to measure open heavy flavour production in proton-proton collisions to provide a normalization for nuclear modifications. Moreover, such measurements allow to test the predictions of theory in the new energy domain. The ALICE apparatus is well suited for detecting open heavy flavour via different channels. First data was collected in 2010 with proton-proton beams at a center-of-mass energy of \unit[7]{TeV}. Promising results were obtained after reconstruction of a subsample of the collected events. $D^0$, $D^+$, $D^{*+}$ and $D_s^+$ mesons were exclusively reconstructed in a wide $p_T$ range. Inclusive spectra of single electrons and muons have been obtained.
\bibliographystyle{aipproc}
\bibliography{General_bib}
\end{document}